\documentclass[12pt]{iopart}

\begin{document}

\title[Aligned neutron--proton pairs in $N\sim Z$ nuclei]
{Aligned neutron--proton pairs in $N\sim Z$ nuclei}

\author{P~Van~Isacker}

\address{Grand Acc\'el\'erateur National d'Ions Lourds, CEA/DSM--CNRS/IN2P3\\
BP~55027, F-14076 Caen Cedex 5, France}
\begin{abstract}
It is argued that $N\sim Z$ nuclei with $90\leq A\leq100$
can be interpreted in terms of aligned neutron--proton pairs
with angular momentum $J=2j$ and isospin $T=0$.
Based on this observation,
a version of the interacting boson model is formulated
in terms of isoscalar high-spin bosons.
To illustrate its possible use,
the model is applied to the $21^+$ isomer in $^{94}$Ag.
\end{abstract}

%Uncomment for PACS numbers title message
%\pacs{00.00, 20.00, 42.10}
% Keywords required only for MST, PB, PMB, PM, JOA, JOB? 
%\vspace{2pc}
%\noindent{\it Keywords}: Article preparation, IOP journals
% Uncomment for Submitted to journal title message
%\submitto{\JPA}
% Comment out if separate title page not required
%\maketitle

\section{Introduction}
\label{s_intro}
One of the goals of radioactive-ion beam facilities
is the uncovering of collective effects
due to isoscalar ($T=0$) neutron--proton (n--p) pairing.
In contrast to the usual isovector ($T=1$) pairing,
where the orbital angular momenta
and the spins of two nucleons are both antiparallel ({\it i.e.}, $L=0$ and $S=0$),
isoscalar pairing requires the spins of the nucleons to be parallel ($S=1$),
resulting in a total angular momentum $J=1$. 
Collective correlation effects conceivably might occur
as a result of isoscalar n--p pairing (and of pairs with $J>1$)
but have resisted so far experimental confirmation~\cite{Warner06}.

Recently, the idea of a pair correlation effect
was proposed by Blomqvist, as described in Ref.~\cite{Cederwall11},
when neutrons and protons are confined to a high-$j$ orbit.
The idea is to interpret the structure of low-energy states of $N\sim Z$ nuclei 
in terms of aligned n--p pairs coupled to maximum angular momentum $2j$.
Currently, $N\sim Z$ experiments are approaching $^{100}$Sn,
and concern nuclei such as $^{92}$Pd~\cite{Cederwall11}
and $^{96}$Cd~\cite{Cederwall11un},
to which Blomqvist's scheme can be applied
since $1g_{9/2}$ supposedly is the dominant orbit in this mass region.

In Ref.~\cite{Zerguine11} Blomqvist's proposal has been examined 
with specific reference to the nuclei $^{96}$Cd, $^{94}$Ag, and $^{92}$Pd,
corresponding to four, six, and eight holes
with respect to the $^{100}$Sn core, respectively.
In addition to the aligned-pair assumption of Blomqvist
the work of Ref.~\cite{Zerguine11} was based on the supplementary hypothesis
that the pairs behave as bosons
and therefore effectively proposed a description of $N\sim Z$ nuclei
in terms of a (non-standard) interacting boson model (IBM)~\cite{Iachello87}.

In this contribution I illustrate,
with the particular example of the $21^+$ isomer in $^{94}$Ag,
how an IBM description in terms of isoscalar high-spin bosons
may elucidate structural issues of $N\sim Z$ nuclei in this mass region.

\section{Aligned isoscalar pairs as bosons}
\label{s_apairs}
Only a summary of results will be given in this section,
referring for full details to Ref.~\cite{Zerguine11}.
The study consisted of two separate parts:
(i) the analysis of shell-model wave functions of $^{96}$Cd
in terms of aligned n--p pairs
and (ii) the mapping of shell-model onto corresponding boson states
for $^{96}$Cd, $^{94}$Ag, and $^{92}$Pd.

(i) For a variety of shell-model interactions appropriate for this mass region,
it was found that the $^{96}$Cd shell-model states 
can be well represented in terms of isoscalar n--p pairs with $J=2j$
(so-called $B$ pairs).
This conclusion came with two caveats.
Firstly, the study~\cite{Zerguine11} did not address (at least not in sufficient detail)
the question whether $1g_{9/2}$ is a dominant orbit in this mass region
but rather {\em presupposed} that it is.
Secondly, the $8^+$ yrast state in $^{96}$Cd
{\em cannot} be written in terms of two $B$ pairs.
This should have observational consequences
in the form of loss of E2 collectivity
between the yrast states of this nucleus.

(ii) An analysis of shell-model eigenstates for more than four nucleons
is a challenging problem which has been studied
by use of the multi-step shell model~\cite{Xu12}.
It is simpler, and at the same time instructive,
to extend the analysis toward higher particle number
through standard boson mapping techniques~\cite{Otsuka78,Skouras90}.
It was found, again with some caveats (for full details see Ref.~\cite{Zerguine11}),
that the complicated spectroscopy
of the nuclei $^{96}$Cd, $^{94}$Ag, and $^{92}$Pd
can, to a large extent, be accounted for
with an interacting boson model
containing a single type of boson with angular momentum $\ell=9$
(a so-called $b$ boson, whence $b$-IBM).

\section{The $21^+$ isomer in $^{94}$Ag}
\label{s_ag94}
Not much is known experimentally about $^{94}$Ag 
except for the presence of two isomers,
with tentative spin-parity assignments $7^+$ (presumably the lowest $T=0$ state)
and $21^+$,
the latter at $6.7(5)$~MeV above the ground state~\cite{Mukha05}.
The shell-model energy of both these states
is reproduced with $b$-IBM to within less than 100~keV~\cite{Zerguine11},
and it can thus be expected
that the latter model provides a good approximation to the former one.
This can be demonstrated explicitly for the $21^+$ isomer,
as I now proceed to show.

In a shell-model description
where three neutrons and three protons
are placed in the $1g_{9/2}$ orbit,
the $21^+$ state is stretched and therefore unique.
In $b$-IBM this state arises from the coupling of three $b$ bosons with $\ell=9$
to total angular momentum $J=21$.
The number of independent states
that can be coupled to total angular momentum $J$
arising from $n$ bosons,
each with individual angular momentum $\ell$,
is given by the sum $\sum_vd^{(\ell)}_v(J)$ (with $v=n,n-2,\dots,1$ or 0)
where the multiplicity $d^{(\ell)}_v(J)$ is known
in terms of an integral over characters
of the orthogonal algebras SO($2\ell+1$) and SO(3)~\cite{Weyl39,Gheorghe04},
\begin{equation}
d^{(\ell)}_v(J)=
\frac{i}{2\pi}
\oint_{|z|=1}
\frac{(z^{2J+1}-1)(z^{2v+2\ell-1}-1)\prod_{k=1}^{2\ell-2}(z^{v+k}-1)}
{z^{\ell v+J+2}\prod_{k=1}^{2\ell-2}(z^{k+1}-1)}dz.
\label{e_mult}
\end{equation}
By virtue of Cauchy's theorem $d^{(\ell)}_v(J)$
is obtained as the negative of the residue of the integrand in~(\ref{e_mult}).
One finds $d^{(9)}_3(21)=2$ and $d^{(9)}_1(21)=0$
and, therefore, two independent boson states with $J=21$
can be constructed,
one of which must be spurious.
As explained in Ref.~\cite{Zerguine11},
the spurious state is eliminated in $b$-IBM
by taking an infinitely repulsive interaction
between two $b$ bosons coupled to angular momentum $J=18$.
Furthermore, since the coefficients of fractional parentage (CFPs)
needed in a three-particle problem are known~\cite{Talmi93},
the following energy expression for the $J=21$ state can be derived:
\begin{equation}
E(b^3;21^+)=3\epsilon_b+
\frac{6851}{20155}\nu^{\rm b}_{12}+
\frac{15488}{21545}\nu^{\rm b}_{14}+
\frac{1212882}{624805}\nu^{\rm b}_{16},
\label{e_e21b3}
\end{equation}
in terms of the two-boson interaction matrix elements
$\nu^{\rm b}_\lambda\equiv\langle b^2;\lambda|\hat V_{\rm b}|b^2;\lambda\rangle$,
and where $\epsilon_b$ is the energy of the $b$ boson.
By virtue of the mapping method,
the boson energy $\epsilon_b$
and the two-boson interaction matrix elements
can be expressed in terms of the shell-model two-body interaction matrix elements
$\nu^{\rm f}_\lambda\equiv\langle(1g_{9/2})^2;\lambda|\hat V_{\rm f}|(1g_{9/2})^2;\lambda\rangle$.
From the mapping of the two-particle system
one finds $\epsilon_b=\nu^{\rm f}_9$.
From the mapping of the four-particle system,
which also can be carried out analytically,
one derives
\begin{eqnarray}
\nu^{\rm b}_{12}&=&
\frac{1218}{69355}\nu^{\rm f}_3+
\frac{63423}{138710}\nu^{\rm f}_4+
\frac{29957}{63050}\nu^{\rm f}_5+
\frac{109881}{53350}\nu^{\rm f}_6+
\frac{1148337}{2358070}\nu^{\rm f}_7
\nonumber\\&&+
\frac{15231}{31525}\nu^{\rm f}_8+
\frac{10893}{535925}\nu^{\rm f}_9,
\nonumber\\
\nu^{\rm b}_{14}&=&
\frac{868}{8515}\nu^{\rm f}_5+
\frac{1953}{1310}\nu^{\rm f}_6+
\frac{46251}{57902}\nu^{\rm f}_7+
\frac{1977}{1310}\nu^{\rm f}_8+
\frac{2211}{22270}\nu^{\rm f}_9,
\nonumber\\
\nu^{\rm b}_{16}&=&
\frac{8}{17}\nu^{\rm f}_7+
3\nu^{\rm f}_8+
\frac{9}{17}\nu^{\rm f}_9.
\end{eqnarray}
By inserting these results into Eq.~(\ref{e_e21b3}),
one finds
\begin{eqnarray}
E_{\rm b}(21^+)&=&
\frac{22134}{3707825}\nu^{\rm f}_3+
\frac{1152549}{7415650}\nu^{\rm f}_4+
\frac{1347751953}{5740387250}\nu^{\rm f}_5+
\frac{8606149749}{4857250750}\nu^{\rm f}_6+
\nonumber\\&&+
\frac{354940047213}{214690483150}\nu^{\rm f}_7+
\frac{1561553973}{220784125}\nu^{\rm f}_8+
\frac{15411107094}{3753330125}\nu^{\rm f}_9.
\label{e_e21f6}
%\\&\approx&
%0.01\nu^{\rm f}_3+
%0.16\nu^{\rm f}_4+
%0.23\nu^{\rm f}_5+
%1.77\nu^{\rm f}_6+
%1.65\nu^{\rm f}_7+
%7.07\nu^{\rm f}_8+
%4.11\nu^{\rm f}_9.
%\nonumber
\end{eqnarray}
This is an approximate expression
since it is derived by use of a boson mapping (whence the index `b').
To what extent it is wrong
therefore yields an idea about the reliability of the boson approximation.

The exact fermionic energy expression
for three neutrons and three protons in a $j=9/2$ orbit,
can be derived with standard techniques involving CFPs~\cite{Talmi93}.
Since the $J=21$ state is unique,
its energy $E_{\rm f}(j^6JT)$ is the matrix element
$\langle j^6JT|\hat V_{\rm f}|j^6JT\rangle=
\sum_\lambda a_\lambda\nu^{\rm f}_\lambda$,
with the coefficients $a_\lambda$ given by
\begin{equation}
a_\lambda=15
\sum_{\alpha'J'T'}
[j^4(\alpha'J'T')j^2(\lambda)JT|\}j^6JT]^2.
\end{equation}
For $j=9/2$, $J=21$ and $T=0$, the following expression results:
\begin{equation}
E_{\rm f}(21^+)=
\frac{21}{65}\nu^{\rm f}_5+
\frac{21}{10}\nu^{\rm f}_6+
\frac{645}{442}\nu^{\rm f}_7+
\frac{69}{10}\nu^{\rm f}_8+
\frac{717}{170}\nu^{\rm f}_9.
\label{e_e21}
\end{equation}
Since the highest allowed angular momentum
for two neutrons and two protons in a $j=9/2$ orbit is $J=16$,
only interaction matrix elements $\nu^{\rm f}_\lambda$ with $\lambda\geq5$
can contribute to the energy of the $J=21$ state in the 3n--3p system.
This rule is obviously obeyed in Eq.~(\ref{e_e21}) but violated in Eq.~(\ref{e_e21f6}).
It is seen, however, that the coefficients of $\nu^{\rm f}_3$ and $\nu^{\rm f}_4$
are rather small in the latter expression,
indicating that the boson approximation is reasonably accurate.

The coefficients $a_\lambda$ in the energy expression
$E_{\rm f}(j^nJT)=\sum_\lambda a_\lambda\nu^{\rm f}_\lambda$
for a unique $n$-particle shell-model state
with angular momentum $J$ and isospin $T$,
satisfy the identities
\begin{eqnarray}
\sum_{\lambda=0}^{2j}a_\lambda=
\frac{n(n-1)}{2},
\nonumber\\
\sum_{\lambda=0}^{2j}\lambda(\lambda+1)a_\lambda=
J(J+1)+j(j+1)\times n(n-2),
\nonumber\\
\sum_{\stackrel{\scriptstyle\lambda=0}{\rm even}}^{2j}2a_\lambda=
T(T+1)+{\frac34}n(n-2).
\end{eqnarray}
These identities are valid for the coefficients in Eq.~(\ref{e_e21}).
It is of interest to note that they 
are also {\em exactly} satisfied by the coefficients in Eq.~(\ref{e_e21f6}).
This reflects the conservation of particle number, angular momentum and isospin
in the shell model,
and the preservation of these quantum numbers under the mapping procedure.

\section{Concluding remark}
\label{s_conc}
More results, for example
concerning the moments of the $21^+$ isomer,
can be derived to test the validity of the boson approximation.
Of more interest will be a similar analysis of the $7^+$ isomer in $^{94}$Ag:
its structure in the shell model, even when confined to the $1g_{9/2}$ orbit,
is complicated with 30 components in the $JT$ scheme
and more than 500 in the $m$ scheme.
In contrast, the number of independent components
is only three in terms of $B$ pairs (or $b$ bosons)
which allows for a better understanding of the structure of the $7^+$ isomer.
A preliminary analysis shows, for example, that its main component
involves two $b$ bosons coupled to intermediate angular momentum 16
which is then coupled with the last boson to total $J=7$.
This problem is currently under further study.

\end{document}